\documentclass[english,doublecol]{epl2} 
\usepackage{babel}
\usepackage[T1]{fontenc}
\usepackage[ansinew]{inputenc}
\usepackage{graphicx}
\usepackage{amsmath}
\usepackage{subfigure}

\newcommand{\Phat}{\hat{P}}
\newcommand{\Rbar}{x}
\newcommand{\zbar}{u}

\newcommand{\ddzz}[1]{\frac{d^2 #1}{d \zbar^2}}

\newcommand{\dt}[3]{\frac{d^{#1} #2}{d #3^{#1}}}

\title{A Separable Solution for the Oscillatory Structure
of Plasma in Accretion Disks}

\author{Massimiliano Lattanzi\inst{1,2} \and Giovanni Montani \inst{1,2,3,4}}
\shortauthor{M. Lattanzi and G. Montani}

\institute{
\inst{1} Department of Physics (G9) - ``Sapienza'' Universit\`a di Roma,
Piazza A. Moro, 5 (00185), Rome, Italy \\
\inst{2} ICRA -- International Center for Relativistic Astrophysics
c/o Dep. of Physics - ``Sapienza'' Universit\`a di Roma\\
\inst{3} ENEA -- C.R. Frascati (Department F.P.N.),
Via Enrico Fermi, 45 (00044), Frascati (Rome), Italy\\
\inst{4} ICRANet -- C.C. Pescara, P. della Repubblica, 10 (65100), Pescara, Italy \\
}


\abstract{
In this paper we provide a new analysis of the system 
of partial differential equations 
describing the radial and vertical equilibria of the plasma 
in accretion disks. In particular, we show that the partial
differential system can be separated once a definite.
oscillatory (or hyperbolic) form for the radial dependence of the relevant
physical quantities is assumed. The system is thus 
reduced to an ordinary differential system in the vertical
dimensionless coordinate. The resulting equations can
be integrated analytically in the limit of small magnetic 
pressure. We complete our analysis with a direct numerical
integration of the more general case. The main result is that a ring-like density profile
(i.e., radial oscillations in the mass density) 
can appear even in the limit of small magnetic pressure.}

\begin{document}

\maketitle

\section{Introduction}

The morphology of an accretion disk around
an astrophysical compact object represents
one of the most important open questions
of stellar physics \cite{B01}.
In fact, while in absence of a significant magnetic
field of the central object the disk configuration
is properly described by the fluid dynamics
approach, the situation becomes puzzling when
we deal with a strongly magnetized source,
namely a pulsar, accreting from a less dense companion.
As recently shown by B. Coppi, see \cite{C05}-\cite{C08},
the plasma nature of the disk implies a significant
coupling between the vertical and the radial
equilibrium, as a consequence of the relevant
Lorentz force acting inside the structure.
The existence of such a coupling suggests a
deep modification of the original point of
view at the base of our understanding of
the stellar accretion phenomenon; for a sample
of the basic literature in the field, see
\cite{PR72}-\cite{RW75}.
In fact, the standard approach to the description
of a thin disk relies on the idea that the
vertical equilibrium can be averaged out when
the viscoresistive MHD is applied to the plasma.
Such a model seems to satisfactorily reproduce the
\emph{coarse-grain} phenomenology, but at the price
of introdocing an anomalous resistivity of the disk
plasma (unjustified by direct estimations),
see Ref. \cite{SS01,Sp08}.\\
The analyses in Refs. \cite{C05,CR06} demostrate that
the details of the disk equilibria are relevant in
establishing an oscillatory local structure inside
the disk. In particular, in Ref.~\cite{CR06} it is shown that,				
for a disk having a sufficiently strong magnetic pressure,
(i. e. a small enough ratio of the thermostatic
pressure to the magnetic one), the mass density
perturbations, due to the internal currents,
are able to induce a ring-like profile.
This ideal MHD result constitutes an opposite
point of view with respect to the idea of a diffusive
magnetic field within the disk, as discussed in Ref. \cite{B01}.
The striking interest in the details of the local
disk morphology consists in the possibilty that
jets of matter and radiation are emitted by virtue
of the strong magnetic field and the axial symmetry,
see for instance Ref. \cite{L96}.

Here, we provide a novel analysis of the fundamental
partial differential system derived in Refs. \cite{C05,CR06}
for the radial and vertical equilibria in the disk plasma.
Our study is based on a separable solution, able
to reduce the coupled partial differential scheme to
a simple ordinary differential system in the vertical
dimensionless coordinate. The separation is
realized by a suitable trigonometric expression for the
radial dependence; the remaining unknowns are
four functions, associated to the vertical
dependence of the magnetic flux function, the mass
density corrections and eventually to functions related to the
different behavior of the thermostatic pressure term.\\
Indeed it is just the presence of two different radial
behaviors of the pressure, the main new feature
of the present analysis. We see that the mass density
perturbations induced in the plasma are relevant
even when the ratio of the thermostatic to the magnetic
pressure is high (differently from the analyis in
\cite{C05}. This correction to the mass distribution
has an oscillating character, so that we see the
formation of the ring-like profile even in the parameter
region where the relevance of the magnetic field is
not crucial on the background level.
This result suggests that the presence of rings in
the disk structure morphology is a very general
feature of the accretion disk structure
for magnetized stars.

The ordinary differential system we derive is analytically integrated
for small values of the ratio of the magnetic pressure over the thermostatic
one; the solution that is obtained is a good tool to fix the
boundary conditions of the numerical analysis for the
general case. In fact, the analytic solution remains
valid for general values of the free parameters, as far as
we restrict ourselves sufficiently close to the equatorial
plane, where the boundary conditions for the numerical
analysis can be given.
The main implication of this link between the analytical
and numerical analyses is that we get a direct
relation between the ratio of the magnetic to thermostatic
pressure to the one between the perturbation wavenumber
and the fundamental wavenumber of the plasma structure.
Then our solution cannot explore the very extreme
value of the parameters, where the magnetic
pressure completely dominates the equilibrium configuration.
The paper is organized as follows. In Sec. I, we describe the 
basic features of the disk. In Sec. II, we write the equations governing 
the radial and vertical equilibrium of the disk. In Sec. III we reduce the
fundamental partial differential system to a system of ordinary differential
equations, and we solve it analytically in the limit of small magnetic pressure, 
discussing the appearance of an oscillatory structure. In Sec. IV
we show the results of the numerical integration, and finally in Sec. V we 
draw our conclusions.

\section{Basic Features of the Disk}

The magnetic field, characterizing the central
object, takes the form 
\begin{equation}
\vec{B} = -\frac{1}{r}\partial _z\psi \vec{e}_r +
\frac{I}{r}\vec{e}_{\phi } + 
\frac{1}{r}\partial _r\psi \vec{e}_z
\, ,
\label{vectorb}
\end{equation}
with $\psi = \psi (r\, ,z^2)$ and
$I=I(\psi \, , z)$.

The matter flux associated with the disk
morphology is:
\begin{equation}
\epsilon \vec{v} =
-\frac{1}{r}\partial _z\Theta \vec{e}_r +
\epsilon \omega (r\, ,z^2)r\vec{e}_{\phi } +
\frac{1}{r}\partial _r\Theta \vec{e}_z
\, ,
\label{solconteq}
\end{equation}
where $\epsilon$ denotes the matter density and
$\Theta (r\, , z)$ is an odd function of $z$, 
to deal with a non zero accretion rate, i.e.
\begin{equation}
\dot{M}_d = -2\pi r\int _{-z_0}^{z_0}\epsilon v_rdz =
4\pi \Theta (r\, , z_0) \equiv 2\pi I 
> 0
\, ,
\label{Mdot}
\end{equation}
$z_0(r)\ll r$ being the half-width of the thin disk.

The similarity of the magnetic field and
matter flux structure, is due to their common divergenceless
nature.\\
Since in the present analysis we are
concentrating our attention on the formation 
of the ring profile within the disk, in what
follows, we neglect the presence of the functions
$I$ and $\Theta$, which are relevant for the
characterization of the azimuthal equilibrium.
In fact, as discussed in \cite{C05,CR06}, the
origin of the oscillatory structure comes out
by the coupling of the vertical and the radial equilibria,
when the internal currents rising in the plasma
are taken into account.

We now develop a local model of the equilibrium,
as settled down around a radius value $r = r_0$,
in order to investigate analytically the effects induced on
the disk profile by the electromagnetic reaction of the plasma.
To this end we split the energy density and the pressure
contributions as
$\epsilon = \bar{\epsilon }(r_0,\, z^2) + \hat{\epsilon}$
and 
$p = \bar{p}(r_0,\, z^2) + \hat{p}$, respectively. 
The same way, we express the magnetic surface function
in the form 
$\psi = \psi _0(r_0) +
\psi _1(r_0\, , r-r_0\, ,z^2)$, with $\psi _1\ll \psi _0$.
The quantities $\hat{\epsilon}$, $\hat{p}$ and $\psi _1$
describe the change of the fundamental
plasma functions due to the currents that emerge 
within the disk embedded into the external magnetic field
of the central object. In general these corrections are
small in amplitude but with a very short scale of variation.
Thus, we are led to address the ''drift ordering'' for
the behavior of the gradient amplitude, i.e.
the first order gradients of the perturbations are of
zero-order, while the second order ones dominate.

As ensured by the corotation theorem
\cite{F37}, the angular
frequency of the disk rotation has to be expressed
via the magnetic flux function as $\omega (\psi )$. 
As a consequence, in the present splitted scheme, 
we can take the decomposition 
$\omega = \omega _K +
\omega ^{\prime }_0\psi _1$,
where $\omega _K$ is the Keplerian term and
$\omega ^{\prime }_0\equiv d\omega_0 d\psi_0 = \mathrm{const}$. This form for $\omega$
holds locally, as far as $(r - r_0)$ remains a sufficiently
small quantity, so that the dominant deviation 
from the Keplerian contribution is due to $\psi _1$.

Accordingly to the drift ordering, the profile of the toroidal
currents rising in the disk, takes the expression $J_{\phi } \simeq -(c/4\pi r_0)\times\left(\partial ^2_r\psi _1 + \partial_z^2\psi _1\right)$.

\section{Vertical and Radial Equilibrium}

We now fix the equations governing the vertical
and the radial equilibrium of the disk, by separating
the basic fluid component from the presence of the
electromagnetic reaction.
Such a splitting of the MHD equations for the
vertical force balance gives
\begin{eqnarray}
\label{verticalequilibrium}
D(z^2) \equiv \frac{\bar{\epsilon}}{\epsilon _0(r_0)} =
e^{-\frac{z^2}{H_0^2}}
\, , \, 
\, H_0^2 \equiv \frac{4K_B\bar{T}}{m_i\omega _K^2}\;,\\
\partial _z\hat{p} + \omega ^2_Kz\hat{\epsilon}
- \frac{1}{4\pi r_0^2}\left(
\partial ^2_z \psi_1 + \partial^2_r\psi _1\right)
\partial _z\psi_1 = 0 
\, ,
\end{eqnarray}
where $\epsilon _0(r_0) \equiv \epsilon (r_0,\, 0)$ and $m_i$ is the ion mass.
The behavior of the function $D(z^2)$
accounts for the pure thermostatic equilibrium
holding in the disk when the vertical gravity
(i. e. the Keplerian rotation) is large enough
to provide a confined thin configuration,
while the temperature $T$ admits the representation
\begin{equation}
2K_BT\equiv m_i\frac{p}{\epsilon} =
m_i\frac{\bar{p} + \hat{p}}{\bar{\epsilon} +
\hat{\epsilon}} \equiv 2K_B(\bar{T} + \hat{T})
\, .
\label{temptot}
\end{equation}
The radial equations underlying the equilibrium
of the rotating layers of the disk, can be decomposed
into the dominant character of the Keplerian angular
velocity plus an equation describing the
behavior of the deviation
$\delta \omega$:
\begin{eqnarray}
\label{radialequilibrium}
2\omega _Kr_0(\bar{\epsilon} + \hat{\epsilon})
\omega _0^{\prime }\psi _1 +
\frac{1}{4\pi r_0^2}\left(
\partial ^2_z \psi_1 + \partial^2_r\psi _1\right)
\partial _r\psi_1 = \nonumber\\
=\partial _r\left[
\hat{p} + \frac{1}{8\pi r_0^2}
\left(\partial_r\psi_1\right)^2\right]
+ \frac{1}{4\pi r_0^2}\partial_r\psi _1 \partial^2_z\psi_1.
\end{eqnarray}
We neglected, in the radial and vertical
equilibria, the presence of the poloidal current
associated with the azimuthal component of the magnetic
field.

We define the dimensionless functions
$Y$, $\hat{D}$ and $\hat{P}$, in place of
$\psi _1$, $\hat{\epsilon}$ and $\hat{p}$, i. e.
\begin{equation}
Y\equiv \frac{k_0\psi _1}{\partial _{r_0}\psi _0}
\, , \,
\hat{D}\equiv \frac{\beta \hat{\epsilon}}{\epsilon _0}
\, , \,
\hat{P}\equiv \beta \frac{\hat{p}}{p_0}
\, ,
\label{deff}
\end{equation}
where $p_0\equiv 2K_B\hat{T}\epsilon _0/m_i$
and $\beta \equiv 8\pi p_0/B^2_{0z} = 
1/(3\epsilon _z^2) \equiv k_0^2H_0^2/3$.
We introduced the fundamental wavenumber $k_0$
of the radial equilibrium, defined as
$k_0\equiv 3\omega _K^2/v_A^2$, with
$v_A^2\equiv 4\pi \epsilon _0/B^2_{z0}$, recalling
that $B_{z0} = \partial _{r_0}\psi _0/r_0$.
It is then natural to deal with the dimensionless
radial variable $x\equiv k_0(r - r_0)$, while
assuming that the fundamental length in the vertical
direction is $\Delta \equiv \sqrt{\epsilon _z}H_0$,
leading to introduce $u\equiv z/\Delta$.1

By this definitions, the vertical and radial equilibrium
equations can be restated respectively as
\begin{eqnarray}
\label{vertad}
\partial _{u^2}\hat{P} + \epsilon _z\hat{D}
+ 2\left(\partial ^2_{x^2}Y +
\epsilon _z\partial ^2_{u^2}Y\right)
\partial _{u^2}Y = 0
\, ,\\[0.3cm]
\label{radad}
\left(D + \frac{1}{\beta }\hat{D}\right) Y + 
\partial ^2_{x^2}Y +
\epsilon _z\partial ^2_{u^2}Y        
+\frac{1}{2}\partial _x\hat{P} + \nonumber \\ 
+\left(\partial ^2_{x^2}Y +
\epsilon _z\partial ^2_{u^2}Y\right)
\partial_xY = 0 
\, .
\end{eqnarray}
Once $D$ and $\hat{D}$ are assigned,
the equations above provide a coupled system 
for $\hat{P}$ and $Y$, allowing to fix the disk
configuration due to the toroidal currents.

\section{Reduction of the Fundamental System}

The analysis of the partial differential system
derived above has been performed in Ref. \cite{C05}
in the limit of small values of $\epsilon _z$ and
an approximated solution was found as an expansion in
such a parameter. Instead in Ref. \cite{CR06}, the study
has been extended to the case $\epsilon _z>1$,
by requiring that the function $Y$ satisfied the
basic eigenstate equation
\begin{equation}
\partial ^2_{x^2}Y +
\epsilon _z\partial ^2_{u^2}Y = - DY  
\, .
\label{radade}
\end{equation}
Here we show that the two partial differential equations
(\ref{vertad}) and (\ref{radad}) can be treated
separating the radial and vertical dependence, thus
reducing them to an ordinary differential system.
In fact, we easily get such a reduction 
by the following positions
\begin{eqnarray}
\label{posfun}
Y = F(u^2)\sin (\alpha x)\\
\hat{P} =  L(u^2)\cos (\alpha x) + M(u^2) \sin ^2 (\alpha x)\\
\hat{D} = d(u^2)\cos (\alpha x)
\, .
\end{eqnarray}
and by imposing the vanishing of the coefficients of
each type of trigonometrical terms.
A simple calculation shows that the 
vertical equilibrium (\ref{vertad})
yields the two equations
\begin{eqnarray}
\label{vertcoup}
\frac{dL}{du^2} + \epsilon _zd = 0\\
\label{vertcoup_2}
\frac{dM}{du^2} + 2\left( -\alpha ^2F + \epsilon _z
\frac{d^2F}{du^2}\right) 
\frac{dF}{du^2} = 0
\, , 
\end{eqnarray}
while the radial equation (\ref{radad}) gives
\begin{eqnarray}
\label{radcoup}
D(u^2)F - \alpha ^2F + \epsilon _z\frac{d^2F}{du^2}
- \frac{\alpha }{2}L = 0\\
\label{radcoup2}
\frac{1}{\beta }dF + \alpha M
+ \alpha F\left( -\alpha ^2F + \epsilon _z
\frac{d^2F}{du^2}\right) = 0 
\, , 
\end{eqnarray}
These two pairs of equations form an ordinary differential
system in the variable $u$ of four coupled second order
equation in the four unknowns $F(u^2)$,
$L(u^2)$, $M(u^2)$ and $d(u^2)$ respectively.
The quantities $\epsilon_z$ and $\alpha $
(we recall that $\beta = 1/3\epsilon ^2_z$)
play the role of free parameters of the problem.
In particular $\epsilon _z$ measures the relevance
of the electromagnetic interaction in the establishment
of the equilibrium configuration of the disk plasma.
The greater $\epsilon _z$ is, the stronger the
internal currents deform the background distribution
of matter and magnetic field.
The parameter $\alpha$ fixes the amplitude of the
radial wavenumber (with respect to the fundamental one $k_0$)
associated to the perturbations. The greater is
$\alpha$, the smaller is the wavelenght of the
radial plasma structures. The same way, also
$\epsilon _z$ can be regarded as the parameter which
gives the scale of the vertcal confinament,
according to the relation, introduced above,
$\Delta = \sqrt{\epsilon _z}H_0$. Such a relation,
together with the definition
$\Delta = \sqrt{H_0/k_0}$, allows to express
the function $D(u^2)$ from $D(z^2)$ introduced in
(\ref{verticalequilibrium}) as
$D(u^2) = \exp \{ -\epsilon _z u^2\}$.

\subsection{Analytical Solution for Small $\epsilon _z$ Values}

Let us study the system of configuration equations
in the limit of small values of the parameter
$\epsilon _z$, 
when we can use the expansion
$D(u^2) = 1 - \epsilon _zu^2$.
In this way, we are led to search a solution to the four equations
above, in the form
\begin{eqnarray}
\label{fsol}
&F = A \exp \left( -\frac{u^2}{2}\right); \;
&L = lF;\\
&d = kF; \;
&M = C(u^2) F^2
\, .
\end{eqnarray}
In other words we assume that the function $Y$ is
confined around the equatorial plane and that the other
functions can be expressed through $F(u^2)$, in agreement
to the structure of the four equations. Substituting expressions (\ref{fsol}) into the equations
(\ref{radcoup}), 
we get the algebraic relations
\begin{eqnarray}
\label{frel}
&\alpha = \frac{1}{2}\sqrt{3(1 - \epsilon _z)}; \;
&l = \frac{2}{3}\alpha;\\
&k = \frac{\alpha }{3\epsilon _z}; \;
&C(u^2) = \alpha ^2 - \epsilon _zu^2
\, .
\end{eqnarray}
Thus, for small values of $\epsilon _z$, we are
able to provide an analytic solution describing the
detailed features of the disk plasma. We see that the wavenumber of the perturbations is not
very different, in this limit, from $k_0$, while
the function $d$ is much greater than $F$ and,
as we shall see below, this is an important peculiar
feature of this solution.

We also remark that, since this solution relies on the expansion
$e^{-\epsilon_z u^2}\simeq 1 - \epsilon_z u^2$, its range of validity
is actually broader than the $\epsilon_z\ll 1$ region. In particular, the solution is still valid
for $\epsilon_z \sim 1$, provided that $u\ll 1/\sqrt{\epsilon_z}$ (i.e., provided
that we are close enough to the equatorial plane). The solution can also 
be continued in the $\epsilon_z >1$ region by noting that in this case,
according to Eq. (\ref{frel}) above, $\alpha$ would be purely immaginary, and
the trigonometric functions would become hyperbolic functions. We are then led to
search a solution in the form:
\begin{eqnarray}
\label{posfun2}
Y' = F'(u^2)\sinh (\alpha' x)\\
\hat{P'} =  L'(u^2)\cosh (\alpha' x) + M'(u^2) \sinh ^2 (\alpha' x)\\
\hat{D'} = d'(u^2)\cosh (\alpha' x)
\, ,
\end{eqnarray}
where, as before:
\begin{eqnarray}
\label{fsol2}
&F' = A' \exp \left( -\frac{u^2}{2}\right) \,; \;
&L' = l'F\,;\\
&d' = k'F \,; \;
&M' = C'(u^2) F'^2
\, .
\end{eqnarray}
Repeating the above procedure we find:
\begin{eqnarray}
\label{frel2}
&\alpha' = \frac{1}{2}\sqrt{3(\epsilon _z-1)} \,; \;
&l' = \frac{2}{3}\alpha'  \, ; \\
&k' = \frac{\alpha' }{3\epsilon _z} \,; \;
&C'(u^2) = -{\alpha'}^2 - \epsilon _zu^2
\, .
\end{eqnarray}

\subsection{The Oscillatory Structure}

Once the form of the solution has been fixed, we can
analyze the physical implications for the disk
structure. In particular, it is immediate to recognize
that for $\epsilon_z < 1$ the mass density distribution acquires
the behavior
\begin{multline}
\frac{\epsilon}{\epsilon_0} = D(u^2) + \frac{1}{\beta} \hat{D}(u^2) = \\
=1 - \epsilon_zu^2 + A(\alpha \epsilon _z)e^{
-\frac{u^2}{2}} \cos (\alpha x) \ge 0
\, .
\label{mdp}
\end{multline}
We see that the perturbations to the mass density profile is an odd function of $x$ and has an
oscillating radial dependence.
On the contrary, it is clear that the oscillating behaviour is not present in 
the $\epsilon_z>1$ regime, since in that case the density and pressure
are expressed in terms of hyperbolic functions. We will then concentrate
in the following in the analysis of the $\epsilon_z<1$ case. \\
Since $\epsilon _z u^2$ is
much smaller than unity, it is easy to realize that the
positive character of the mass density is ensured by
the request $A < 1/(\alpha \epsilon_z)$. 
The total pressure term in the disk plasma
is given by
\begin{multline}
\frac{p}{p_0} = e^{ -\epsilon _zu^2}
+ 3\epsilon _z^2A\bigg[ \frac{2}{3}\alpha \cos (\alpha x) + \\
A\left( \alpha ^2 - \epsilon _zu^2\right)
e^{- \frac{u^2}{2}} \sin ^2(\alpha x) \bigg]
e^{ -\frac{u^2}{2}}
\, .
\label{precon}
\end{multline}
and the perturbation is an even function of $x$. 
The term in squared brackets can provide a negative
contribution to the total pressure, but, in the limit
of small enough $\epsilon _z\ll 1$, its weight is
limited by the coefficient $\epsilon _z^2$ and
the expression above, for values of $u$ in
the range of some units over the equatorial plane,
rewrites as 
\begin{equation}
\frac{p}{p_0} \simeq 1 + 3\epsilon _z^2A
\left[ \frac{1}{\sqrt{3}}\cos (\alpha x) +
\frac{3}{2}A
e^{ - \frac{u^2}{2}} \sin ^2(\alpha x) \right]
e^{ -\frac{u^2}{2}}
\,  
\label{precon1}
\end{equation}
and the total plasma pressure is clearly positive
when $A < 1/ (\alpha \epsilon _z^2)$. It can be seen
that, since $\epsilon_z <1$, this is always ensured
once the condition for the positiveness of
the density ($A < 1/ (\alpha \epsilon _z)$ ) is fulfilled.
However, in the general case, the positive character
of $p/p_0$ in the point $\alpha x = \pi$,
requires that $2\alpha \epsilon _z^2A\exp [(\epsilon _z - 1/2)u^2] \ll 1$,
that constrains the possible values of $A$,
with implications on the morphology of the ring
profile.

\begin{figure}[th]
\begin{center}
\subfigure{
\includegraphics[clip, width=0.35\textwidth]{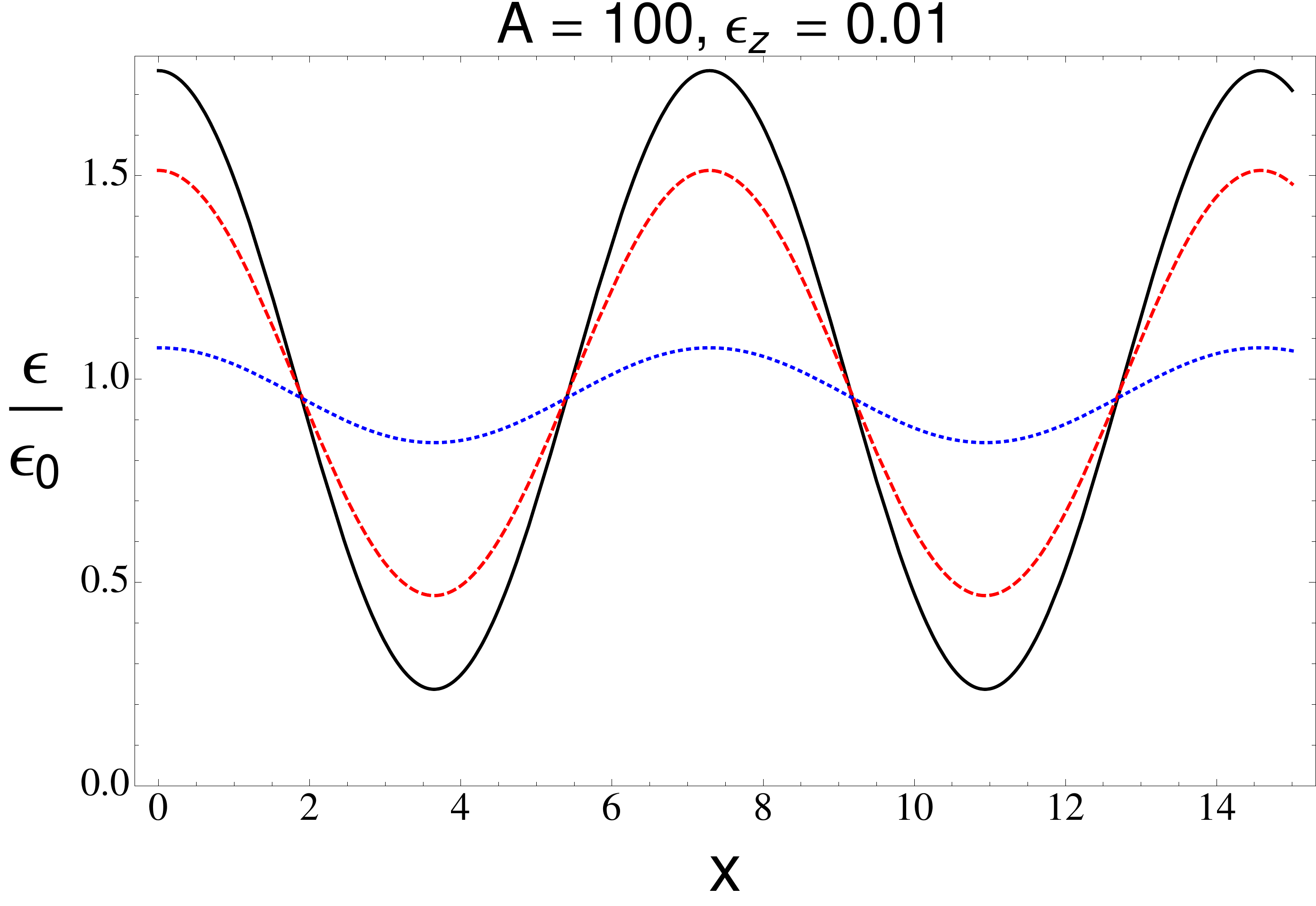}
}
\hspace{0.5cm}
\subfigure{
\includegraphics[clip, width=0.35\textwidth]{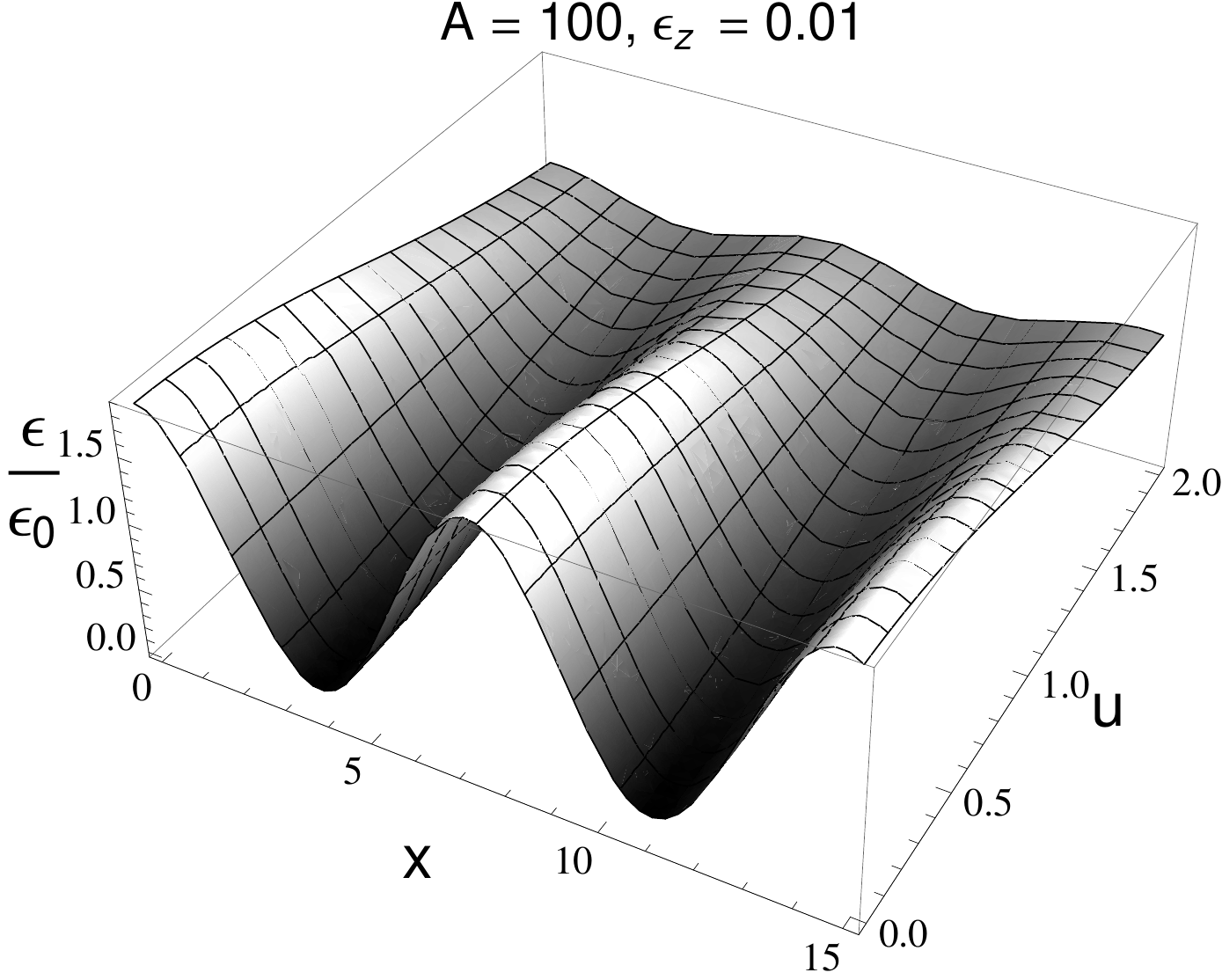}
}
\caption{Top: Radial density profile of the disk for $u = 0.5, 1, 2$ (black [solid], red [dashed] and blue [dotted] curves, respectively). Bottom:
Normalized mass density as a function of the dimensionless radial and vertical coordinates $x$ and $u$. In both panels, $A=100$ and $\epsilon_z=0.01$.}
\label{fig:dens}
\end{center}
\end{figure}

\begin{figure}[th]
\begin{center}
\subfigure{
\includegraphics[clip, width=0.35\textwidth]{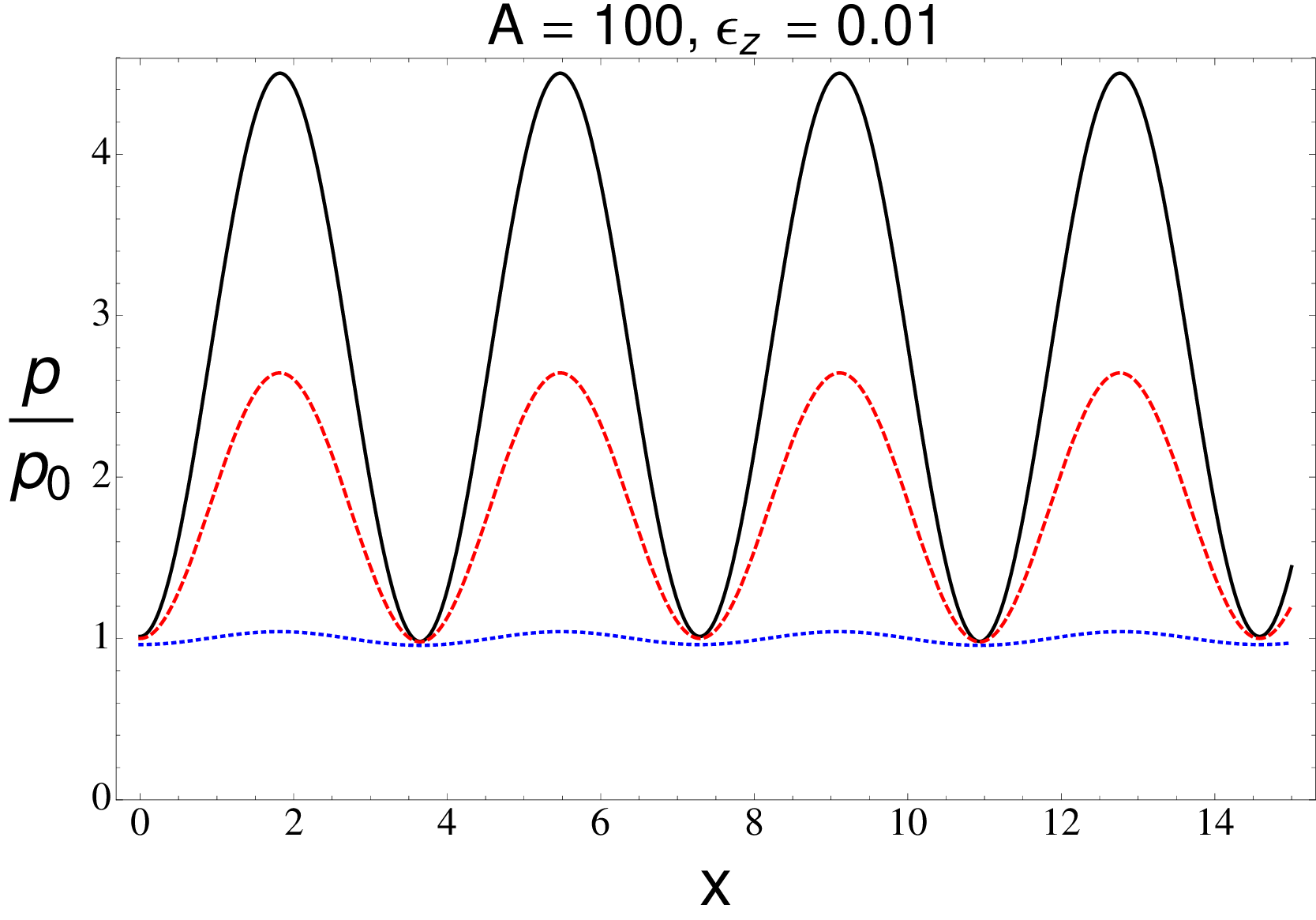}
}
\hspace{0.5cm}
\subfigure{
\includegraphics[clip, width=0.35\textwidth]{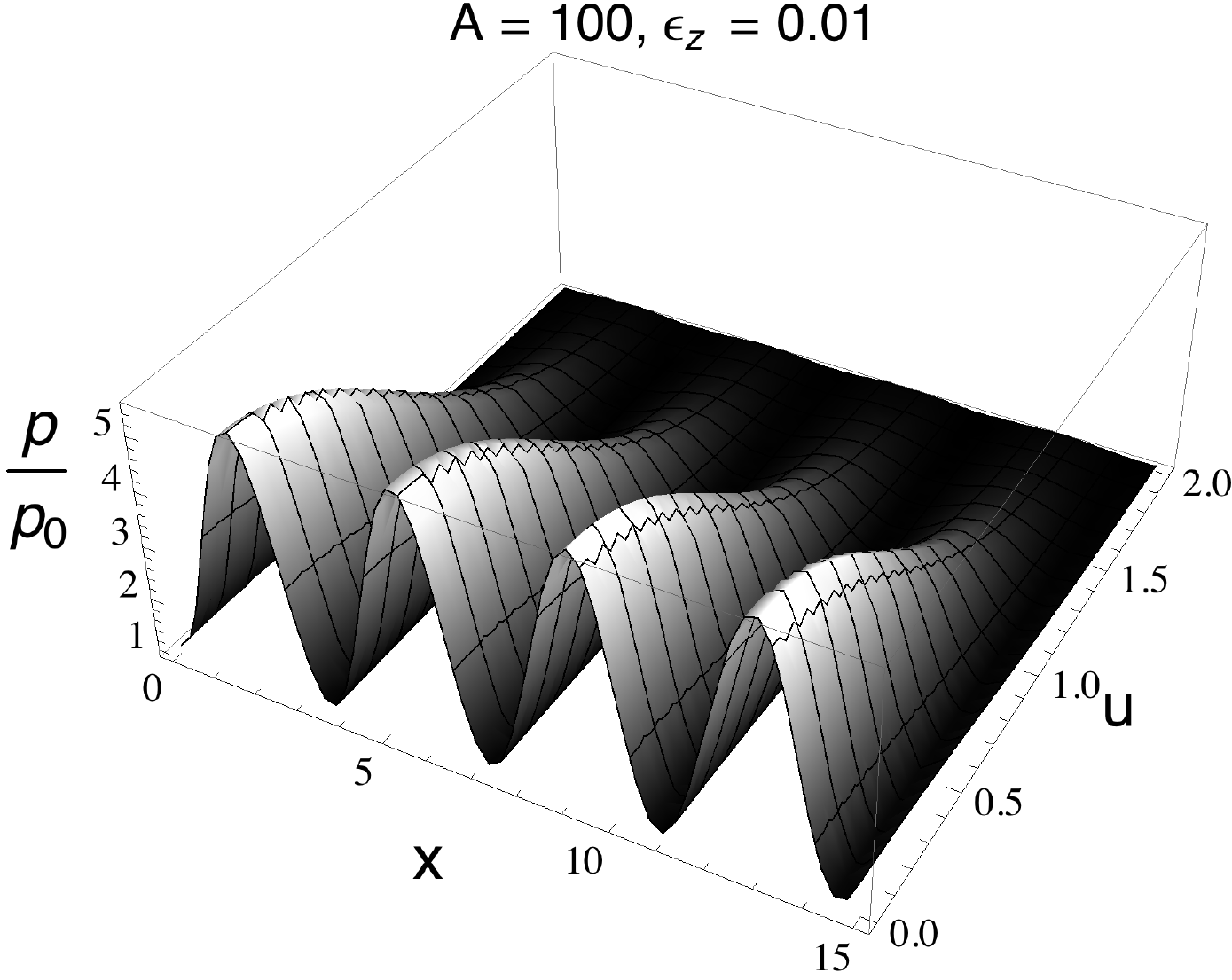}
}
\caption{Top: Normalized pressure as a function of the dimensionless radial coordinate $x$, for $u = 0.5, 1, 2$ (black [solid], red [dashed] and blue [dotted] curves, respectively).
Bottom: Normalized pressure as a function of the dimensionless radial and vertical coordinates $x$ and $u$.
In both panels, $A=100$ and $\epsilon_z=0.01$.}
\label{fig:press}
\end{center}
\end{figure}

However, the very important feature we get, is that,
differently from the analysis in Refs.  \cite{C05,CR06},
here the mass density can have nodes even for small
values of $\epsilon_z$, i. e. for high $\beta $ values
of the plasma (see Fig. \ref{fig:dens}). This behavior is a
consequence of dealing with a solution in which
the quantity $\hat{\epsilon}$ is of order $\epsilon _z$,
instead of order $\epsilon_z^2$ like in \cite{C05,CR06}.\\
We show the radial density and pressure profiles in Figs. \ref{fig:dens} and \ref{fig:press}, 
for different values of the vertical coordinate $u$,
and for values of the parameters $A=100$ and $\epsilon_z=10^{-2}$
[the value of $\alpha$ is fixed by the relation (\ref{frel})]. It is clearly
seen that there are ``empty'' regions, i.e. regions where $\epsilon/\epsilon_0 \ll 1$,
indicating the presence of a ring-like structure of the disk. 
This a general feature of the mass distribution described by Eq. (\ref{mdp}) 
when $A \alpha \epsilon_z \simeq 1$. 
The relevance of this result relies on the existence
of a local ring profile in the disk plasma even
if the magnetic pressure does not dominate the termostatic one.
As a consequence, we can infer that the oscillatory
structure of the disk is expectedly a very diffuse
phenomenon in accreting astrophysical sources. 

\emph{Comparison with previous works}

In this section we compare our results with those obtained in previous works, mainly focusing
on Ref. \cite{CR06}. The main assumption underlying the analysis presented in Ref. \cite{CR06} is that
$Y$ satisfies the eigenvalue equation (\ref{radade}). In our analysis, this equation is not satisfied given the positions that we have made concerning the form the functions $Y$ and $\Phat$
[see Eq. (\ref{posfun}) above]. 
In fact, substituting $Y~=~F(\zbar^2) \sin{(\alpha \Rbar)}$ in the above eigenvalue equation, we get:
\begin{equation}
\epsilon_z \dt{2}{F}{\zbar} -\alpha^2 F+\bar D F=0.
\end{equation}
On the other hand, we have that instead, in our analysis, the corresponding equation satisfied by
$F$ is Eq. (\ref{radcoup}), namely:
\begin{equation}
\epsilon_z  \ddzz{F}  - \alpha^2 F +\bar{D} F - \frac{\alpha}{2} L  = 0.
\end{equation}
We recall that the function $L(\zbar^2)$ appearing in the additional term 
is related to the cosine part of the pressure.
It it is thus clear that, if $L(\zbar^2)\ne 0$, the two approaches will bear different results. 
So the basic difference between the two analyses can be traced back to the presence,
in the pressure function, of a term proportional to $\cos{(\alpha \Rbar)}$. 

\section{Numerical Analysis}

In the region of the parameters where
$\epsilon _z$ is not much smaller than unity,
the system (\ref{vertcoup})-(\ref{radcoup}) has to be
integrated numerically. To this purpose,
we change the vertical variable to 
$\xi = u^2/2$. Observing that
$\frac{d^2F}{du^2} = 2\xi \frac{d^2F}{d\xi ^2} +
\frac{dF}{d\xi} $
we rewrite the configuration system in the form:
\begin{eqnarray}
\label{vertcoup2}
\frac{dL}{d\xi} + 2\epsilon _zd = 0\\
\frac{dM}{d\xi} + 2\left[ -\alpha ^2F + \epsilon _z\left( 
2\xi \frac{d^2F}{d\xi ^2} +
\frac{dF}{d\xi}\right)
\right]
\frac{dF}{d\xi} = 0
\, . \\
\label{radcoup_2}
e^{ -2\epsilon _z\xi}F - \alpha ^2F +
\epsilon _z\left(
2\xi \frac{d^2F}{d\xi ^2} + \frac{dF}{d\xi}\right) 
- \frac{\alpha }{2}L = 0\\
3\epsilon _z^2
dF + \alpha M
+ \alpha F\left[ -\alpha ^2F +
\epsilon _z \left( 
2\xi \frac{d^2F}{d\xi ^2} + \frac{dF}{d\xi}\right) \right] = 0
\, .
\end{eqnarray}
This system of four ordinary differential equations in the four 
unknown functions $F(\xi),\,L(\xi),\,M(\xi)$ and $d(\xi)$ can 
be integrated numerically by fairly standard methods. 
However, in order to perform the numerical integration,
initial conditions should be given. These can be found
by noting that, when $\xi$ is small enough, the analytic
solution found in the previous section is an acceptable
solution to the differential system even if $\epsilon_z$ 
is of order unity or larger. In fact, the parameter of the 
expansion that leads to the above analytical solution
is $\epsilon_z u^2 = 2\epsilon_z \xi$, so that
for every value of $\epsilon_z$
it is always possible to find a (however small) starting
value $\xi_0$ for the numerical integration such that $2\epsilon_z \xi_0 \ll 1$.

\begin{figure}[ht]
\begin{center}
\includegraphics[clip, width=0.4\textwidth]{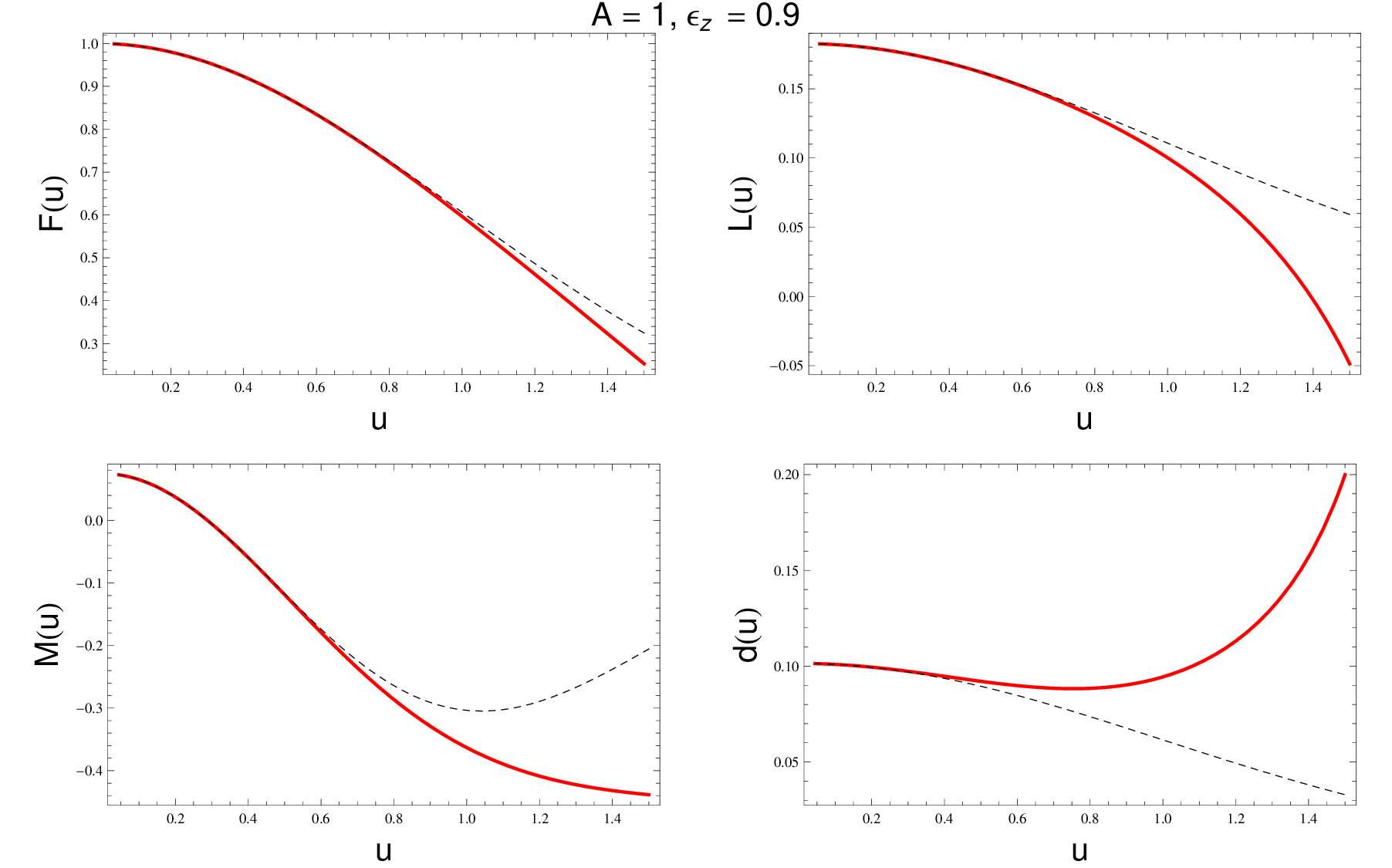}
\caption{Behaviour of the functions $F(u)$, $L(u)$, $M(u)$ and $d(u)$ for $D(u)=e^{\epsilon_z u^2}$, $A=1$ and $\epsilon_z=0.9$ (Red [solid] lines). Also shown is the
correspoding analytical solution for $D(u)=1-\epsilon_z u^2$ (Black [dotted] lines).}
\label{fig:num1}
\end{center}
\vspace{-0.5cm}
\end{figure}

\begin{figure}[ht]
\begin{center}
\includegraphics[clip, width=0.4\textwidth]{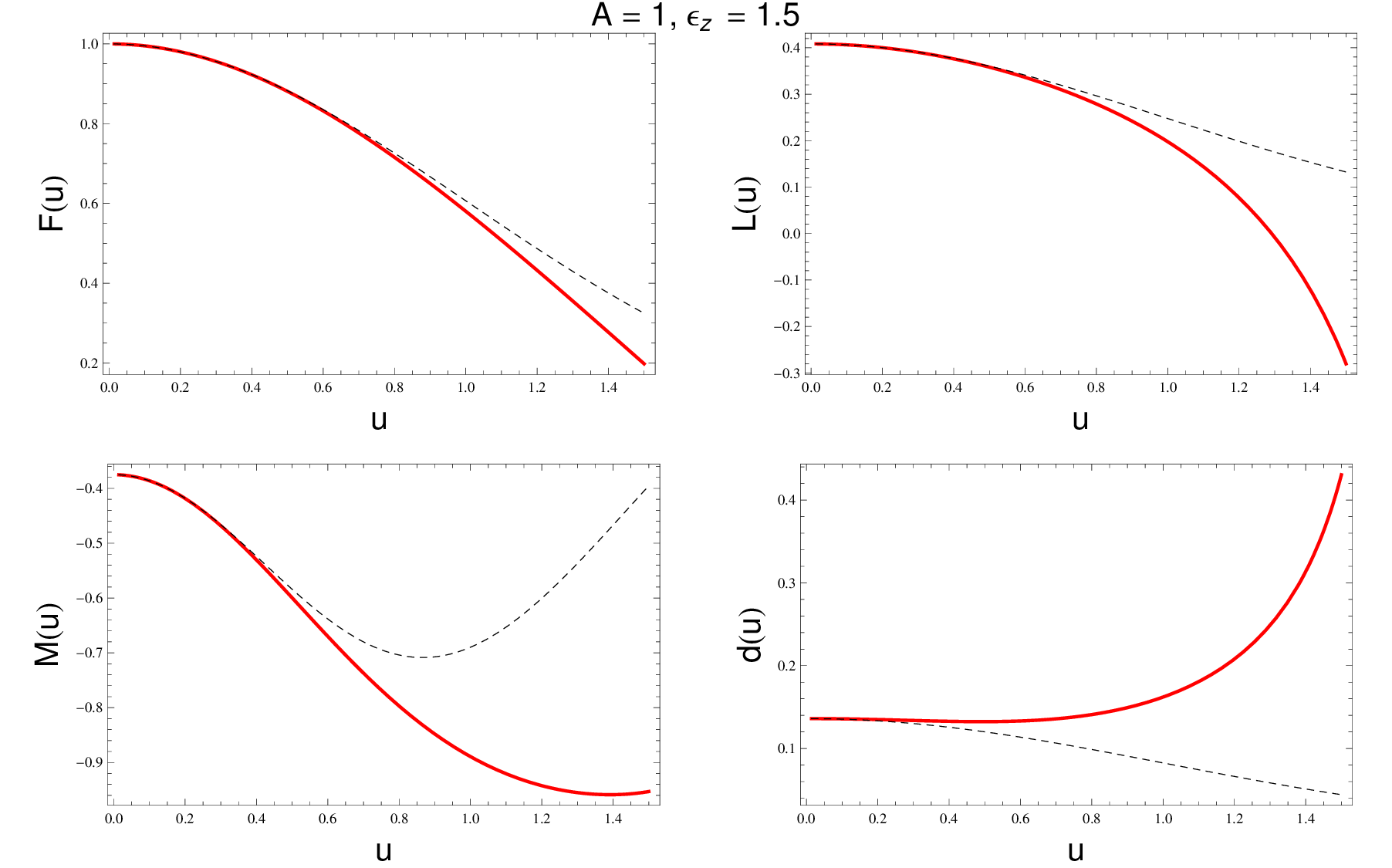}
\caption{The same as Fig. \ref{fig:num2}, but for $\epsilon_z=1.5$}.
\label{fig:num2}
\end{center}
\vspace{-0.5cm}
\end{figure}

Then, we give the initial conditions so that they match the analytic solution in $\xi_0$:
\begin{eqnarray}
\label{cond}
F_0\equiv F(\xi_0) = A e^{-\xi_0}\,; \quad
L(\xi_0) = \frac{2}{3}\alpha F_0\,; \nonumber \\
d(\xi_0) = \frac{\alpha}{3\epsilon _z} F_0;\quad
M(\xi_0) = (\alpha ^2-2\epsilon_z\xi_0) F_0^2; \nonumber\\
\frac{dF}{d\xi }(\xi_0) = - F_0
\end{eqnarray}
The fact that we are matching the analytical solution in $\xi_0$ means that we have to
enforce the condition $\alpha=\displaystyle\frac{1}{2}\sqrt{3(1-\epsilon_z)}$ like we did before,
in order to ensure the consistency of the differential system.
Hence, initial conditions are completely fixed once $A$ and $\epsilon_z$ (or equivalently $\alpha$) are fixed.
The results of our numerical integration are shown in Figs. \ref{fig:num1}-\ref{fig:num2}.

\section{Concluding Remarks}
In this paper we presented an exact separable
solution of the radial and vertical equilibrium
equations at the basis of the oscillating morphology
emerging in the plasma configuration of an
accretion disk. We reduced the original system
of two partial differential equations into a
set of four independent ordinary differential
equations in the vertical coordinate,
one for the flux surface function, one for the
mass density perturbations and two
related to different components of the pressure.
This reduced system admits an analytical
solution only in the limit of high values of
the $\beta$ parameter of the plasma, i. e.
only when the thermostatic pressure of the disk
is much larger than the magnetic pressure
contribution. We also performed a numerical 
analysis of the system for small values of $\beta$.

The main results of our analysis
can be summarized in the following three
points.

i)-We have derived a solution showing
how the radial gradient of the thermostatic
pressure is relevant in establishing the
equilibrium, even in the linear limit and for small
values of $\epsilon _z$. This feature makes
our solution intrinsically different from the
analyses developed in Refs. \cite{C05,CR06}, except
for the discussion of the extreme non-linear
regime in Section IX of Ref. \cite{CR06}.
In fact, in such a limit, our approach is
reconcilied with that one, because the latter
accounts on an equivalent level the role of the
radial gradient of the pressure.

ii)-We have obtained the fundamental feature that the ring-like
structure emerges as a strong structural feature
of the plasma disk confined in magnetic field,
since the radial oscillation of the mass density
takes place even in the linear weakly magnetized
limit $A\ll 1$ and $\epsilon _z\ll 1$. Indeed a suitable
choice of these free parameters of the model is always
possible in order to arrange for nodes in the
mass density profile. This morphology is relevant because
suggests that the ring profile can be expected to
be a general character of the magnetized accreting
compact objects we observe in the Universe,
and it is due to the direct link existing between
the radial pressure gradient and the mass density
perturbations. This output of our separation
algorithm indicates that more general configuration
scenarios can be contained in the radial and
vertical equilibrium equation with respect to
the one investigated in \cite{C05,CR06}, though
they could exist in a non-separable regime.

iii)-We have found that the disk profile undergoes a
transition when the parameter $\epsilon _z$
becomes greater than one, going from the oscillating structure
to an hyperbolic behavior. This fact suggests the
possible co-existence of two different disk components
in the same global profile. In fact in the
present local model we addressed $\epsilon _z$
as a constant because it refers to a
generic value of the radial coordinate $r_0$, but throught
the disck it is clearly a function of the radial coordinate itself,
taking values above and below unity in different radial regions.

\section{Acknowledgments}

We would like to thank Riccardo Benini for his valuable advice.
This work has been developed in the framework of the CGW collaboration
(www.cgwcollaboration.it).

\end{document}